\documentclass[]{spie}  

\usepackage{amsmath,amsfonts,amssymb,xspace}
\usepackage{graphicx}
\usepackage[colorlinks=true, allcolors=blue]{hyperref}

\newcommand{\mum}{\mbox{{\usefont{U}{eur}{m}{n}{\char22}}m}\xspace}

\title{AO3k at Subaru: First on-sky results of the facility extreme-AO}

\author[a]{Julien Lozi}
\author[b]{Kyohoon Ahn}
\author[a]{Hannah Blue}
\author[c]{Alicia Chun}
\author[d]{Christophe Clergeon}
\author[a]{Vincent Deo}
\author[a,e,f,g]{Olivier Guyon}
\author[a]{Takashi Hattori}
\author[a]{Yosuke Minowa}
\author[h]{Shogo Nishiyama}
\author[a]{Yoshito Ono}
\author[a]{Shin Oya}
\author[a]{Yuhei Takagi}
\author[a,g]{S\'{e}bastien Vievard}
\author[i]{Maria Vincent}
\affil[a]{Subaru Telescope, National Astronomical Observatory of Japan, National Institutes of Natural Sciences (NINS), 650 North A`oh\={o}k\={u} Place, Hilo, HI 96720, United States}
\affil[b]{Korea Astronomy and Space Science Institute Daedeokdae-ro 776, Yuseong-gu Daejeon 34055, Republic of Korea}
\affil[c]{University of Chicago, 5801 S. Ellis Ave., Chicago, IL 60637, United States}
\affil[d]{Gemini International Observatory, a program of NSF's NOIRlab, 670 North A`oh\={o}k\={u} Place, Hilo, HI 96720, United States}
\affil[e]{Steward Observatory, University of Arizona, Tucson, AZ 87521, United States}
\affil[f]{College of Optical Sciences, University of Arizona, Tucson, AZ 87521, United States}
\affil[g]{Astrobiology Center of NINS, 2 Chome-21-1, Osawa, Mitaka, Tokyo, 181-8588, Japan}
\affil[h]{Miyagi University of Education, 149, Aramaki-aza-Aoba, Aobaku, Sendai}
\affil[i]{University of Hawai`i Institute for Astronomy, 2680 Woodlawn Dr, Honolulu, HI 96822, United States}

\authorinfo{Further author information: (Send correspondence to J.L.)\\J.L.: E-mail: lozi@naoj.org, Telephone: 1 808 934 5949}

\begin{document} 
\maketitle

\begin{abstract}
The facility adaptive optics of the Subaru Telescope AO188 recently received some long-awaited upgrades: a new 3224-actuator deformable mirror (DM) from ALPAO (hence the name change to AO3000 or AO3k), an upgraded GPU-based real-time computer, a visible nonlinear curvature wavefront sensor and a near-infrared wavefront sensor (NIR WFS), closing the loop at up to 2~kHz. The wavefront sensors were added in 2023, while the DM will be installed at the beginning of 2024. With these new features, AO3k will provide extreme-AO level of correction to all the instruments on the IR Nasmyth platform: The NIR-MIR camera and spectrograph IRCS, the high-resolution Doppler spectrograph IRD, and the high-contrast instrument SCExAO. AO3k will also support laser tomography (LTAO), delivering high Strehl ratio imaging with large sky coverage. 

The NIR WFS, using part of the light from y- to H-band, is dramatically increasing the number of reachable targets for high-contrast imaging (HCI), for exoplanets characterization, as well as AGNs or the galactic center. It has two modes that can be used to drive the new DM: A double roof-prism pyramid WFS, and a focal plane WFS for faint targets.

The high Strehl will especially benefit SCExAO for high-contrast imaging, both in infrared and visible. The second stage extreme AO will no longer have to chase large residual atmospheric turbulence, and will focus on truly high-contrast techniques to create and stabilize dark holes, as well as coherent differential imaging techniques. We will finally be able to leverage the several high performance coronagraphs tested in SCExAO, even in the visible.

AO3k will answer crucial questions as a precursor for future adaptive optics systems for ELTs, especially as a technology demonstrator for the HCI Planetary Systems Imager on the Thirty Meter Telescope. A lot of questions are still unanswered on the on-sky behavior of high actuator counts DMs, NIR wavefront sensing, the effect of rolling shutters or persistence.

We present here the first on-sky results of AO3k, before the system gets fully offered to the observers in the second half of 2024. These results give us some insight on the great scientific results we hope to achieve in the future.

\end{abstract}

\keywords{extreme adaptive optics, pyramid wavefront sensor, focal plane wavefront sensor, near infrared wavefront sensor, deformable mirror, nonlinear curvature wavefront sensor}

\section{INTRODUCTION}
\label{sec:intro}  

The first generations of adaptive optics (AO) and extreme AO (XAO) are getting upgraded with a larger number of actuators, faster low-noise detectors, more sensitive wavefront sensors (WFS) and even new near-infrared WFS (NIR WFS).

At the Subaru Telescope, following the success of the first AO system AO36, AO188 was developed and installed in 2006\cite{Takami2006}, including a laser guide star\cite{Hayano2006}. With 188 actuators in the pupil, AO188 can provide Strehl ratios of $\sim20$--40\% in H-band in median seeing\cite{Minowa2010}. Three instruments are using the output of AO188:
\begin{itemize}
    \item The facility instrument IRCS (Infrared Camera and Spectrograph), providing both imaging (y- to M'-bands) and spectroscopy (zJ- to L-band) capabilities.
    \item The PI XAO platform SCExAO (Subaru Coronagraphic Extreme Adaptive Optics), feeding several science modules in visible (R- and i-band) and NIR (y- to K-band), including the integral field spectrograph CHARIS, the MKID Exoplanet Camera (MEC) and the newly-upgraded visible differential polarimetric imager VAMPIRES\cite{Jovanovic2015,Lozi2018,Lucas2024}.
    \item the PI instrument IRD (Infrared Doppler spectrograph), a NIR (y- to H-band) fiber-fed high resolution spectrograph. IRD can be fed directly with multi-mode fibers from AO188, or with single-mode fibers from SCExAO (REACH module)\cite{Kotani2018}.
\end{itemize}

In the broader context of Subaru 2\footnote{Subaru 2 website: \url{https://subarutelescope.org/jp/subaru2/}}, aiming at enhancing Subaru's functionalities in key aspects, such as wide-field, high-resolution observations and infrared astronomy, we want to push for better and faster wavefront control by upgrading components of AO188, SCExAO, and the whole instrument architecture of the infrared Nasmyth platform of Subaru. Finally, we want to test technologies necessary to prepare for the next generation of giant segmented telescopes like the Thirty Meter Telescope (TMT)\cite{Ono2020}. The first steps aim at upgrading AO188 into AO3k,an XAO platform combining ALPAO's 3224-actuator deformable mirror (later referred as DM3k), a near infrared wavefront sensor (NIR WFS) using a double roof-prism pyramid already commissioned on-sky\cite{Lozi2022,Lozi2023} and a non-linear curvature wavefront sensor (nlCWFS)\cite{Ahn2024}.

In this paper, we will first present the steps envisioned to upgrade the facility adaptive optics and the instrument configuration (Sec.~\ref{sec:phases}). Then we will detail journey to get the DM3k, its laboratory characterization and installation inside the instrument (Sec.~\ref{sec:dmjourney}). We will show the first on-sky results obtained with the new DM3k combined with the NIR WFS, as well as the first demonstration of two XAO systems, AO3k and SCExAO, running at the same time (Sec.~\ref{sec:ao3k_onsky}). Finally, we will conclude and give some perspective to the work ahead of us.

\section{AO3\MakeLowercase{k}: AN UPGRADE OF AO188 IN PHASES}
\label{sec:phases}

The current facility adaptive optics of the Subaru Telescope AO188 is composed of a 188-actuator bimorph DM from Cilas (named after DM188), coupled with a visible curvature wavefront sensor (CWFS) using avalanche photo-diodes (APDs). A vibrating membrane modulate the signal at 2~kHz, So the loop runs at 1~kHz. It provides Strehl ratios between 20 and 40\% in H-band in median seeing. 

Higher performances are now becoming essential to improve the scientific output of the instruments using AO, by reaching deeper contrasts for high-contrast imagers, reaching fainter targets for imaging and spectroscopy, and reaching redder targets (M-type stars, stars embedded in dust, e.g. with protoplanetary rings, or around the galactic center)

The goal is to upgrade the facility adaptive optics in phases, with minimal impact on the observing schedule and current functionalities.

\begin{itemize}
    \item \textbf{Phase Ia: Addition of a NIR WFS}.
    
    The goal of this phase was to add a NIR WFS inside the existing enclosure, without affecting the current AO188 performances and functionalities. It was integrated inside AO188 for the first time in May 2022, and tested on-sky during some engineering time in May 2022 and May 2023. In 2022, we only operated the NIR WFS in a passive way, while closing the loop with the current visible CWFS. In 2023, we performed the first closed-loop tests using the 188-actuator bimorph DM. At the same time, a new real-time controller (RTC) was developed to be able to control the NIR WFS, as well as the older hardware, e.g. the DM188 and the APDs.

   \begin{figure}
   \begin{center}
   \begin{tabular}{c}
   \includegraphics[width=0.97\textwidth]{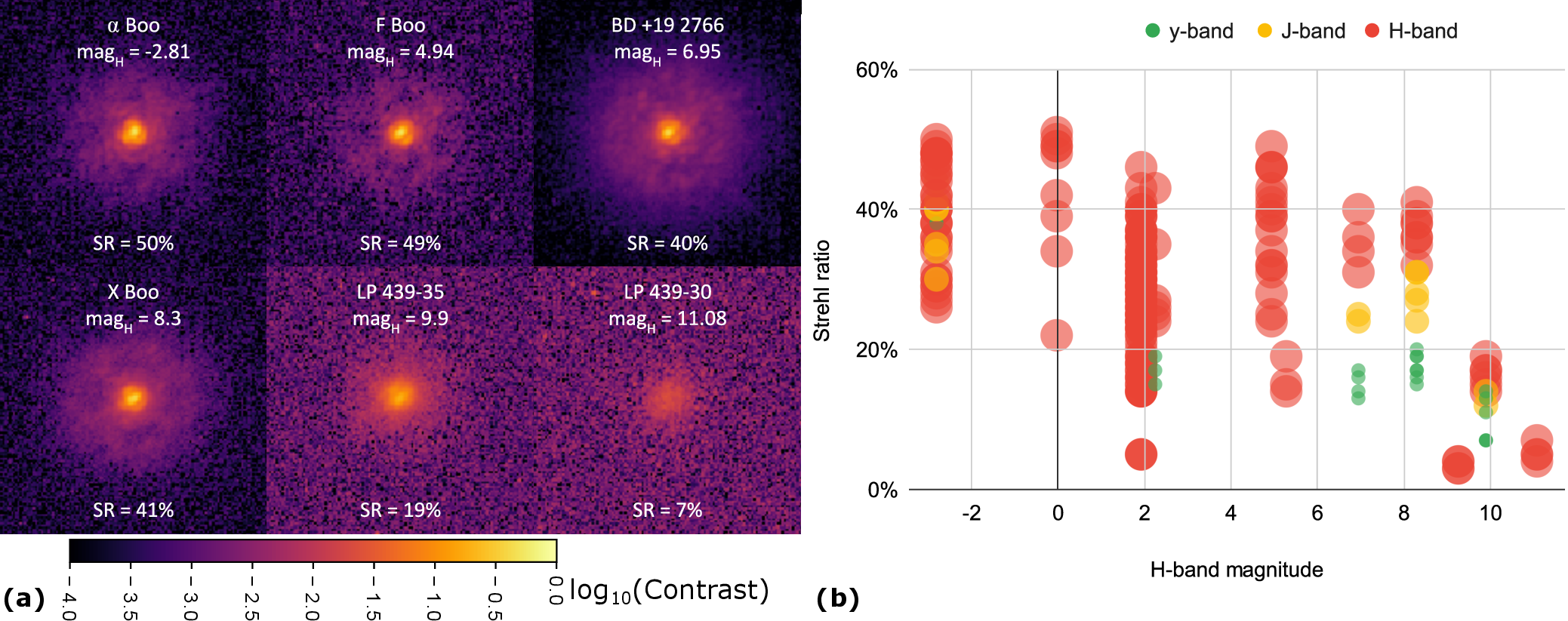}
   \end{tabular}
   \end{center}
   \caption{(a) H-band Strehl ratio measurements obtained with SCExAO's Palila camera, over a few seconds, for targets with various magnitudes. Each image is 2x2". (b) All Strehl ratio measurements taken with Palila in y-, J- and H-band.}
   \label{fig:srmeasurements} 
   \end{figure}

    Figure~\ref{fig:srmeasurements} (a) presents a sample of H-band images taken with SCExAO's Palila camera (First Light Imaging C-RED2, named after a critically endangered Hawaiian honeycreeper found on the slopes of Maunakea) during the May 2023 engineering run, to measure the Strehl ratio, for several targets with various magnitudes. The targets were chosen in the same part of the sky to have similar atmospheric conditions for each of them. The airmass was between 1 and 1.4 for all of them, although the seeing was variable.

    Figure~\ref{fig:srmeasurements} (b) shows a compilation of all the Strehl ratio measurements taken with Palila in y-, J- and H-band during the engineering run. This includes times where we were tuning various parameters, so some Strehl values can be much lower than optimal. The higher values for each magnitude would represent the performance of the loop in normal conditions. The Strehl ratio is over 40\% until magnitude ~9, then decreases.

    During the engineering time, we demonstrated we could use the NIR WFS with the DM188 and achieve similar and even better performance than with the CWFS. Thanks to these on-sky tests, the NIR WFS is now available for open-use observations since S24A, and Phase Ia is completed.

    \item \textbf{Phase Ib: Replacement of the original 188-actuator DM (DM188) with ALPAO’s 64x64 DM (DM3k)}.
    
    In this phase, the number of available actuators in the pupil will increase from 188 to $\sim3000$. Therefore, AO1888 will be re-branded as AO3k. Once the DM is replaced, the priority is to demonstrate that we can maintain the current performance of AO188 by down-sampling the DM and using the current APDs for wavefront control. It ensures that we do not need to suspend Open-Use observations and do a full characterization of the system before offering it for science. We can do this characterization over time, once all the elements are in place. The XAO performance of AO3k is achieved using the NIR WFS added in Phase Ia and is the main focus of this paper.
    
    The new DM3k is a drop-in replacement of the original one, although the DM188 was also mounted in a fast tip/tilt mount (TTM) (100~Hz) where offloads of these modes are sent. To maintain this offload capability in this phase, the fast TTM is equipped with a flat mirror and moved to the fold mirror right before the DM (AOM2 mirror). Since the tip/tilt motion will not be in a pupil plane anymore, the largest offsets are offloaded to the telescope pointing more regularly than before (typically every tens of seconds instead of every tens of minutes).
    
    In parallel, we developed a  non-linear CWFS to replace the original CWFS, this time using a sCMOS camera and imaging simultaneously four out-of-focus pupil images. This new nlCWFS is capable of controlling all the modes offered by the DM3k and will provide the full XAO performances of AO3k in the future. This new wavefront sensor will be tested on-sky with the DM3k in August 2024.\cite{Ahn2024}
    
    \item \textbf{Phase II: Installation of a NasIR beam switcher for up to 4 instruments}.
    
    A Nasmyth IR beam switcher (NBS) was designed to switch rapidly between up to 4~instruments without requiring craning as it is currently done behind AO188/AO3k.\cite{Hattori2024} The light can potentially be split between 2~instruments (e.g. SCExAO and IRCS) with a dichroic beamsplitter.
    
    In this phase, the NIR WFS will move out of AO3k, on a common platform with the laser tomography AO (LTAO) currently in development\cite{Akiyama2020}, a precursor of the future MCAO ULTIMATE-SUBARU. The NBS is in its integration phase, and will be added behind AO3k early 2025.

    Two new instruments will be added thanks to the NBS: SPIDERS\cite{Thompson2024}, a high-contrast high-resolution spectro-imaging test platform, and NINJA\cite{Tokoku2022}, a multi-purpose visible-NIR slit spectrograph.
\end{itemize}

\section{ALPAO'S DM3\MakeLowercase{k}: A JOURNEY}
\label{sec:dmjourney}

\subsection{There and Back Again}
\label{sec:history}

ALPAO built a first 3224-actuator DM for ESO in 2018, as a demonstrator for larger versions for ELT\cite{Vidal2019}. The same design was the perfect size as a replacement for the DM188, so we started its procurement between 2019 and 2021, after securing the funding through various grants. Due to the time between the manufacture of two DMs, ALPAO struggled to meet our specs, and for contractual reasons, delivered a product with known issues in March 2022. Even if this first version did not performed as we hoped  (missing actuators, rest shape that could not be flatten with the stroke), we still kept it for a few months, to develop our testbed in Hilo, Hawai`i, and our testing software. The first version was sent back to France in August 2022.

The ALPAO team ended up rebuilding the DM head several times and found similar issues, leading them to change their fabrication process and quality control. In addition, post-covid supply chain issues compounded on the delays.

Towards the end of 2023, the final version was successfully built and tested. It was completed in December, and delivered in January 2024. When we plugged it the first time, we noticed what looked like a bad actuator that was not present before shipping. After investigation, we found out more issues, most notably a strong gradient in the gain map and 5-6x less stroke than expected. After some back-and-forth with the ALPAO team, on-site repairs were scheduled in March 2024.

The ALPAO engineers concluded that all the issues were linked to the same root cause, the shifting of an internal reference plate during shipping. They successfully completed the repairs on March 17, and we started the characterization right after.

\subsection{Laboratory Characterization}
\label{sec:lab}

The characterization of the DM3k was perform in Subaru's base facility in Hilo, Hawai`i, on a crunched timeline due to all the delays, since the on-sky engineering schedule, planned months in advance, did not change.

In a clean booth, we tested the DM3k on its mount, facing a Zygo interferometer. Most of the characterization was done with the Zygo, although we did some dynamical tests with the nlCWFS in parallel\cite{Ahn2024}.

\begin{figure}
\begin{center}
\begin{tabular}{c}
\includegraphics[width=0.97\textwidth]{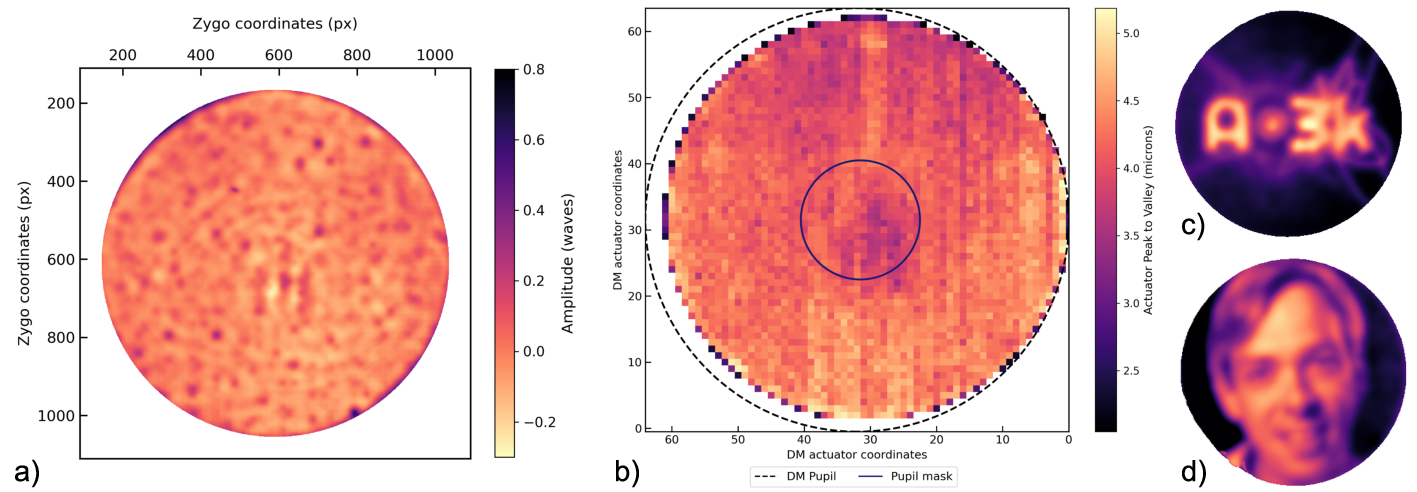}
\end{tabular}
\end{center}
\caption{(a) Flatness of the DM after several flattening iterations with the Zygo. The RMS value is 31~nm, although the edge of the DM is cropped here. (b) Gain map for every actuator, using the Zygo. A small up/down gradient is observed, as well as some weaker actuators close to the center. The small circle represents the central obscuration, masking most of the weak actuators. (c) and (d) AO3k logo and Olivier's picture applied to the DM and measured with the Zygo, showing the resolution and the DM.}
\label{fig:dm_characterization} 
\end{figure} 

The Zygo was used to flatten the DM iteratively. Figure~\ref{fig:dm_characterization} (a) presents the result of such a process. The residual wavefront error measured here is 31~nm, although we managed to go down below 20~nm on other occasions. The edges of the mirror were cropped in this measurement, because they curl in such a way that is not measurable by the Zygo. This effect impacts slightly the final pupil edge, and we will discuss this later.

A gain map was also computed by poking each actuator individually. The result is presented in Figure~\ref{fig:dm_characterization} (b). Once again, edge actuators are omitted because they could not be measured. The final gain map shows a slight up/down gradient, and a region with smaller gain (named the "lazy" actuators), close to the center. The circle at the center represents the obscuration of the telescope M2 mirror, which masks most of the lazy region. So they should not impact the final performance. The stroke was measured to be up to specs, to about 40~\mum (wavefront) for a 3x3 region.

Figure~\ref{fig:dm_characterization} (c) \& (d) are patterns that were applied to the DM (the AO3k logo and SCExAO's PI Olivier Guyon respectively), showcasing that even if the influence function blurs the image a little bit, the actuator density is impressive.

\begin{figure}
\begin{center}
\begin{tabular}{c}
\includegraphics[width=0.97\textwidth]{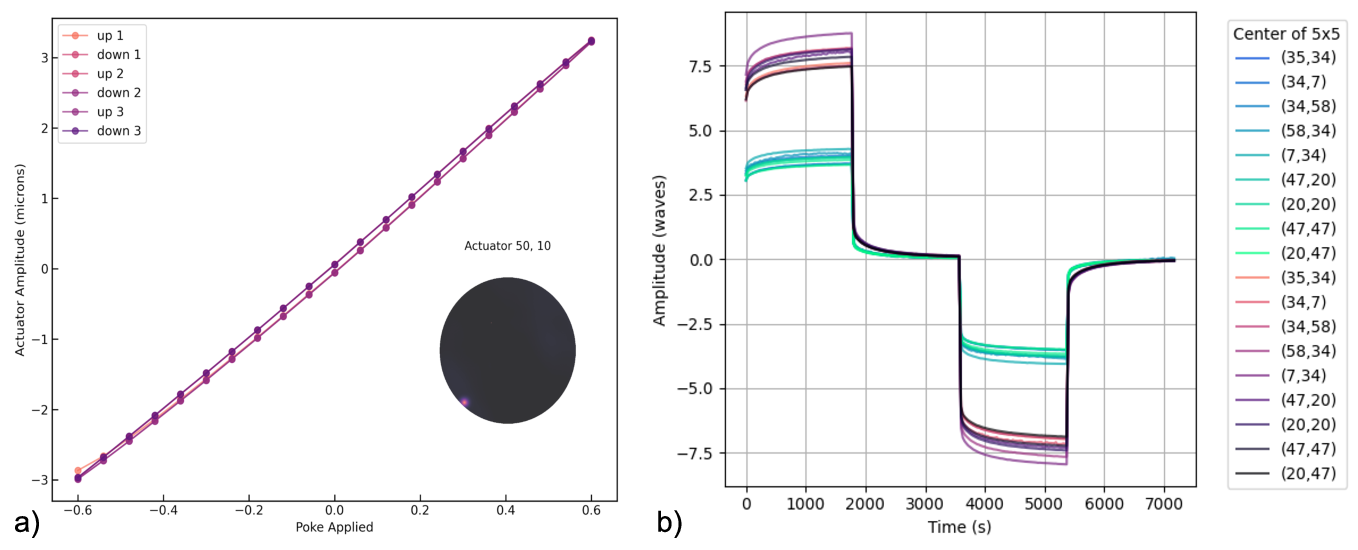}
\end{tabular}
\end{center}
\caption{(a) Linearity and hysteresis measurement for one actuator. The linearity is good and hysteresis in the order of a few percents. All actuators have very similar profiles. (b) Measurement of the settling time (or creep) for different regions. The DM will take approx. 30 minutes to settle to its final amplitude, when the command is static. It is proportional to the amplitude.}
\label{fig:dm_linearity} 
\end{figure} 

We also looked at the linearity and hysteresis, by applying a ramp of commands on individual actuators and repeating the process several times. An example is presented in Figure~\ref{fig:dm_linearity} (a), although all the actuators tested have an almost identical profile. Both linearity and hysteresis are shown to be on the order of a few percents, and the repeatability is at least an order of magnitude better.

From the results of the first 3224-actuator DM\cite{Vidal2019}, we know that the DM3k has a "warm-up" period, i.e. actuators will continue to drift for a while after a command is applied. This behavior is shown in Fig.~\ref{fig:dm_linearity} (b), where several 5x5 regions on the DM were poked positively, then back to zero after 30~minutes, then negatively after 30~minutes, then poked negatively after another 30~minutes, and finally back to zero again. We can see that every time, the motion takes about 30~minutes to settle, and drifts for about 20\% of the original amplitude independently of the poke amplitude. In a dynamic closed-loop environment such as an AO system correcting atmospheric turbulence, this behavior is not an issue, since we are constantly adjusting the shape of the mirror. But it will play a role in a more static configuration, for example if we want to apply a flat map. For example, the flat map calculated for Fig.~\ref{fig:dm_characterization} (a) will drift and not provide a flat wavefront for very long. This would be a "cold" flat map, to be used right away after being applied. It is possible to compute a "warm" flat map though, that would take into account this drift. In that case, the DM would not be flat at first, but would slowly drift to a flat wavefront after 30~minutes.

After these tests, the DM3k was deemed up to specs, and we were authorized to go ahead with the upgrade of the instrument. Some contingency plans to roll back to the original configuration were still put in place if the installation was unsuccessful before the next science observation.

\subsection{Installation at the Telescope and Preliminary Testing}
\label{sec:installation}

\begin{figure}
\begin{center}
\begin{tabular}{c}
\includegraphics[width=0.97\textwidth]{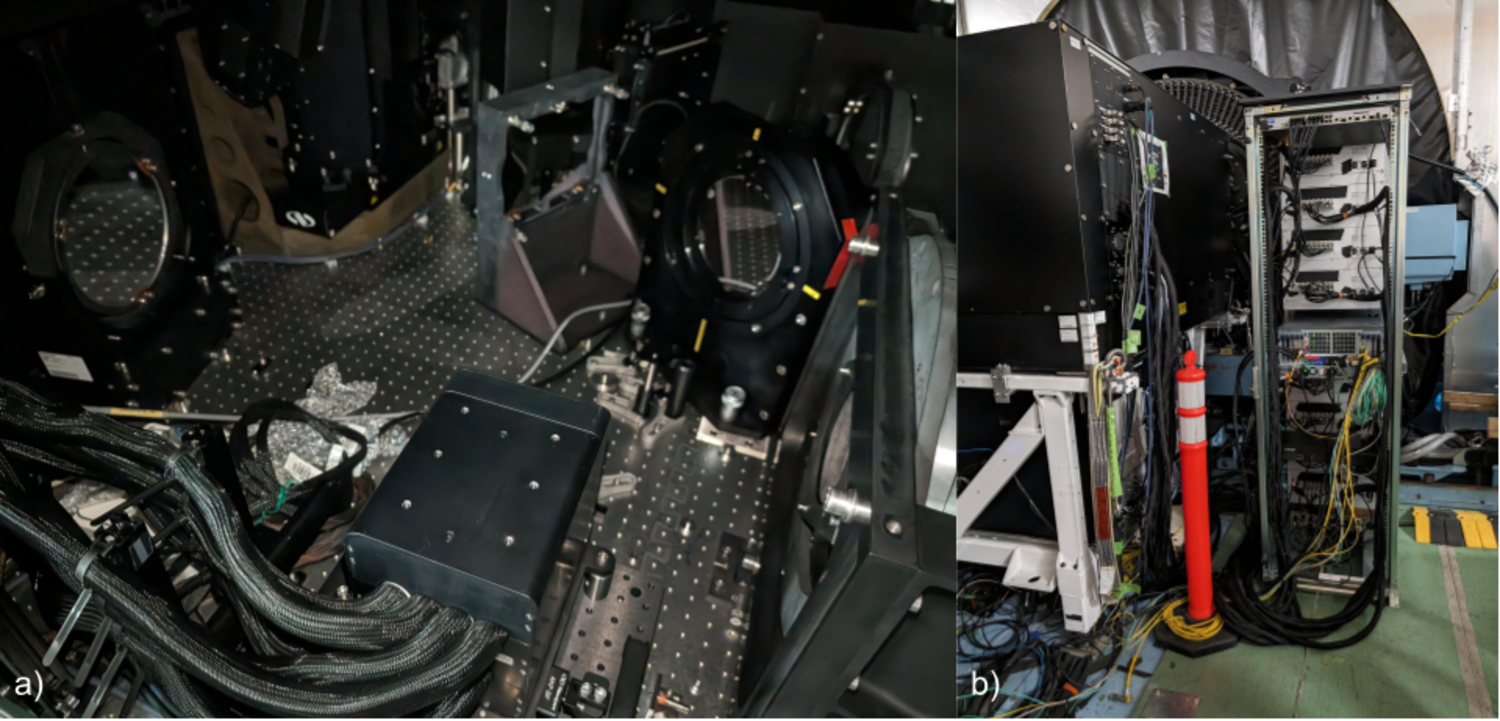}
\end{tabular}
\end{center}
\caption{(a) The DM3k installed inside AO3k (foreground). The fast TTM of the original DM of AO188 was moved to the flat mirror AOM2 right before the DM in the optical path (background, the mount with red and yellow tapes). (b) The DM electronics and the new RTC are populating a full rack located next to the instrument, due to DM cable lengths.}
\label{fig:dm_installation} 
\end{figure}

The installation at the summit started on May 3, 2024, and was completed after 10 days. We started by removing the DM188 from the TTM, and measured the TTM response for increasing mass simulators (700~g to 1.4~kg), since the AOM2 mirror is about 2~times heavier than the DM188. The response did not change significantly from the DM188, so we proceeded with the installation of the AOM2 mirror and the migration of the mount in the location of AOM2. Then the DM3k was installed in place of the DM188 and aligned. Figure~\ref{fig:dm_installation} (a) shows the inside of AO3k after installation. The new DM3k with its cables can be seen in the foreground, and the TTM with AOM2 can be seen in the background. The new RTC and the 8~DM electronics are populating a rack located right next to the instrument (Figure~\ref{fig:dm_installation} (b). Right now, the heat from the rack (approx. 3~kW) is just dumped on the Nasmyth platform and managed by the air conditioning, but we are considering upgrading to a liquid-cooled rack in the future.

\begin{figure}
\begin{center}
\begin{tabular}{c}
\includegraphics[width=0.97\textwidth]{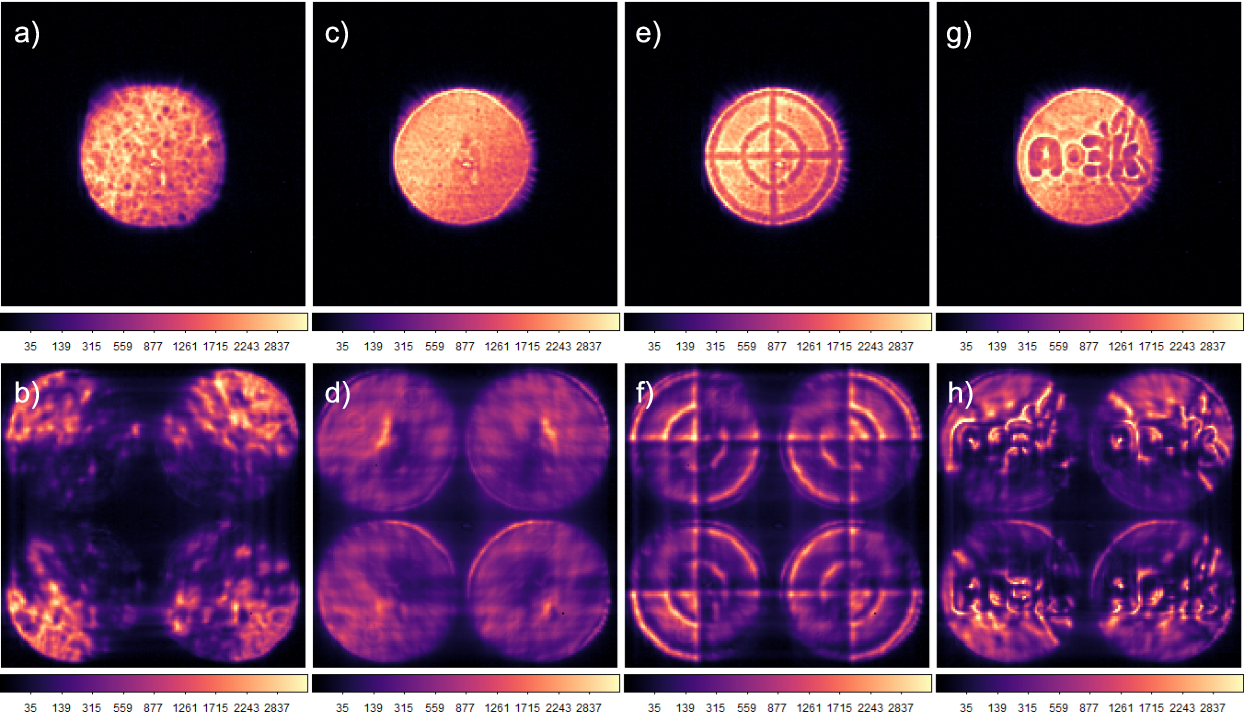}
\end{tabular}
\end{center}
\caption{NIR WFS near-pupil images (top row) and modulated PyWFS images (bottom row) of: (a) \& (b) the DM in its rest shape, (c) \& (d) the flat map calculated in the lab with the Zygo, (e) \& (f) a bullseye pattern used for alignment, (g) \& (h) the AO3k logo applied to the DM.}
\label{fig:nirwfs_patterns} 
\end{figure}

Once the DM3k was in place and aligned, we performed the first tests using the internal lasers (633~nm and 1550~nm) and the wavefront sensors (visible CWFS and NIR WFS). We first checked that the DM was responding as well as in the lab, and that it was not damaged by the transport to the summit. Figure~\ref{fig:nirwfs_patterns} presents an example of tests that were performed. Here we applied various patterns and looked at the result on the NIR WFS. Two modes were used: a near-pupil plane imaging mode (top row), and the modulated PyWFS mode (bottom row). Four patterns are presented on this figure, the rest shape (voltage zero), the flat map calculated with the Zygo in the lab, a bullseye pattern (added to the flat map) for alignment of the DM surface with the beam, and the AO3k logo (added to the flat map again). Since the lab flat map is working well, we confirmed that the DM was working as expected and we did not have last minute surprises.

Once noticeable thing though is the fuzziness of the edge especially in pupil plane imaging mode. We can see that the curliness of the surface at the edge is affecting the pupil. Indeed, the reflective surface is 96.5~mm in diameter, while the beam is 90x94~mm due to the incidence angle of $16^o$. This effect is less pronounced in visible, probably due to the shorter wavelength. One side is more affected than the other, which led us to translating the DM vertically in relationship to the beam. This is possible because we do not need a perfect registration of the actuators in relationship with the WFS pixels, since it is calibrated with the response matrix.

The new RTC used to control the DM3k uses CACAO\cite{Guyon2018,Deo2024}. The RTC can also control the original DM, the TTM, as well as the wavefront sensors: The APDs of the CWFS, the new nlCWFS, and the NIR WFS. It is also equipped with a pass-through mode, designed to allow us to keep using the original RTC (deployed in 2006), and keep all the original functionalities (e.g. the LGS mode) without forcing us the develop updated versions all at once. 

\begin{figure}
\begin{center}
\begin{tabular}{c}
\includegraphics[width=0.6\textwidth]{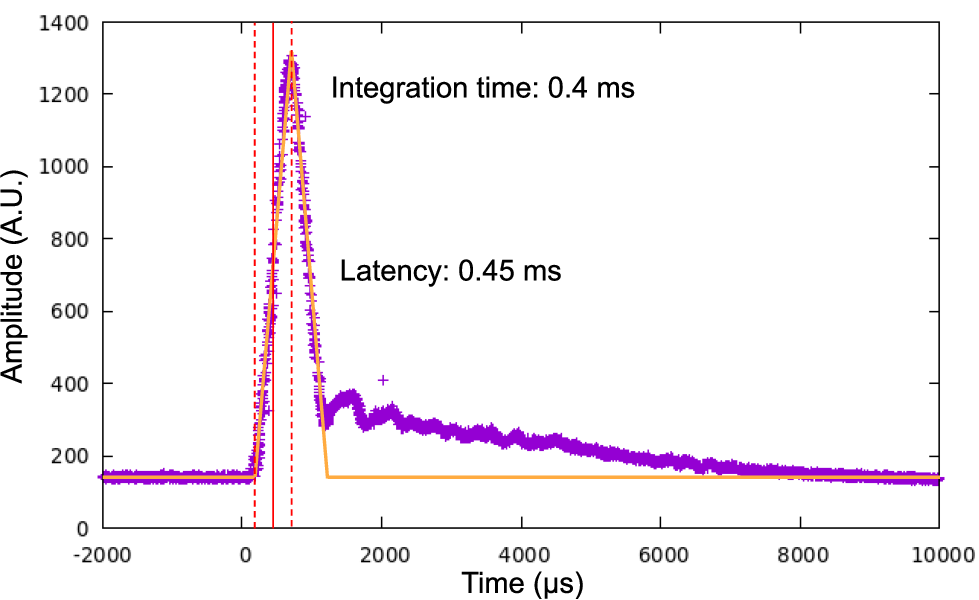}
\end{tabular}
\end{center}
\caption{Measurement of the hardware latency with CACAO, using the NIRWFS as the wavefront sensor. The orange triangular profile is the ideal case, with the rising time corresponding to the exposure time of the WFS, and the halfway rising point the hardware latency, independent of the loop frequency. The trailing end corresponds to mechanical resonances of the DM surfaces settling in about 10~ms.}
\label{fig:dm_latency} 
\end{figure} 

One of the functionalities of CACAO is to calculate with high precision the hardware latency for any control loop (the "mlat" function). Figure~\ref{fig:dm_latency} presents a measurement, using the NIR WFS at 2~kHz, and the DM3k. The orange triangular shape is the ideal output. The rising time of the triangle corresponds to the integration time of the WFS, while the half-way point of that same rising slope corresponds to the hardware latency. This latency is independent from the loop frequency, and only take into account the reaction time of the DM, the communication delays, and the readout time of the WFS. This latency is measured at 0.45~ms in this case. The trailing end is due to mechanical resonances on the DM surface that create ripples when the DM is poked. It takes about 10~ms for these ripples to settle. This behavior was known since the first DM was tested. This will probably have an impact on the final performance of the AO system, but this DM behaves way better than the DM188 in this regard (see latency measurement at 2~kHz in \cite{Lozi2023}). 

The operation of the AO system was validated with the new system and the internal source, and we proceeded with the first on-sky tests with two half nights on May 23 and 24, 2024, with SCExAO behind AO3k, and one full night on May 26 with IRCS. Here we present results with SCExAO, since it can provide the most telemetry. The seeing during these nights were good to excellent (0.6" down to 0.25"!).

\section{FIRST ON-SKY RESULTS OF AO3\MakeLowercase{k}}
\label{sec:ao3k_onsky}

\subsection{Validation of AO188 Mode for Uninterrupted Open-Use Observations}
\label{sec:ao188mode}

\begin{figure}
\begin{center}
\begin{tabular}{c}
\includegraphics[width=0.97\textwidth]{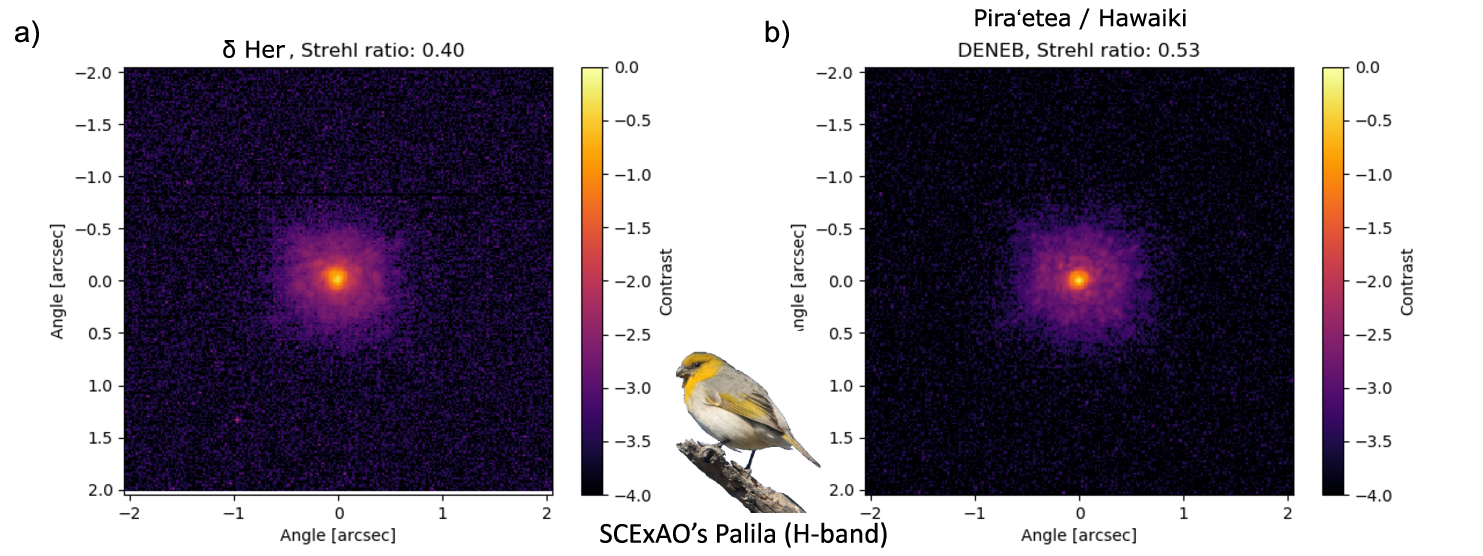}
\end{tabular}
\end{center}
\caption{AO188 mode performance validation with H-band Strehl ratio measurements. They were obtained with SCExAO's Palila camera, over a few seconds, on the target (a) $\delta$ Her and (b) Pira`etea/Hawaiki (Deneb).}
\label{fig:palila_ao188} 
\end{figure} 

The highest priority was to make sure that the performance with in the AO188 mode (i.e. using the original CWFS) were not impacted by the change of DM. This was paramount, since regular science observations were continuing a few weeks later, on June 14. Therefore we tested the control of the DM3k using the orignal CWFS, both in pass-through mode (using the original RTC), and with the new RTC. 

Figure~\ref{fig:palila_ao188} presents two examples of Strehl ratio measurements using SCExAO's Palila camera, in H-band. In these examples, only the AO188 loop is used, while SCExAO is used passively to image the Point Spread Function (PSF). Figure~\ref{fig:palila_ao188} (a) presents a Strehl measurement of 40\% on the star $\delta$ Her, while Fig.~\ref{fig:palila_ao188} (b) shows a Strehl of 53\% on Pira`etea or Hawaiki (respectively a traditional name of the star Deneb in the Society Islands, French Polynesia, and a modern Hawaiian name for the same star). Thanks to a more stable DM, we can achieve better performance with the same wavefront sensor than before with AO188. 

These results confirmed that changing the DM will have no negative impact on the AO correction for the rest of the scheduled science nights. 

\subsection{First Results of Two XAO Loops Running Together}
\label{sec:2XAO}

For the first time ever, two XAO systems, AO3k ---with the NIR WFS using the First Light Imaging C-RED ONE running at 1~kHz, and ALPAO's DM3k--- and SCExAO ---with the visible PyWFS using the First Light Imaging OCAM2k running at 2~kHz, and Boston Micromachines Corporation (BMC) MEMS 2k DM--- were used at the same time to correct the atmospheric turbulence. with such a number of actuators and such speeds, it was sometimes hard to distinguish what we were seeing from laboratory PSFs with the calibration source.

\begin{figure}
\begin{center}
\begin{tabular}{c}
\includegraphics[width=0.97\textwidth]{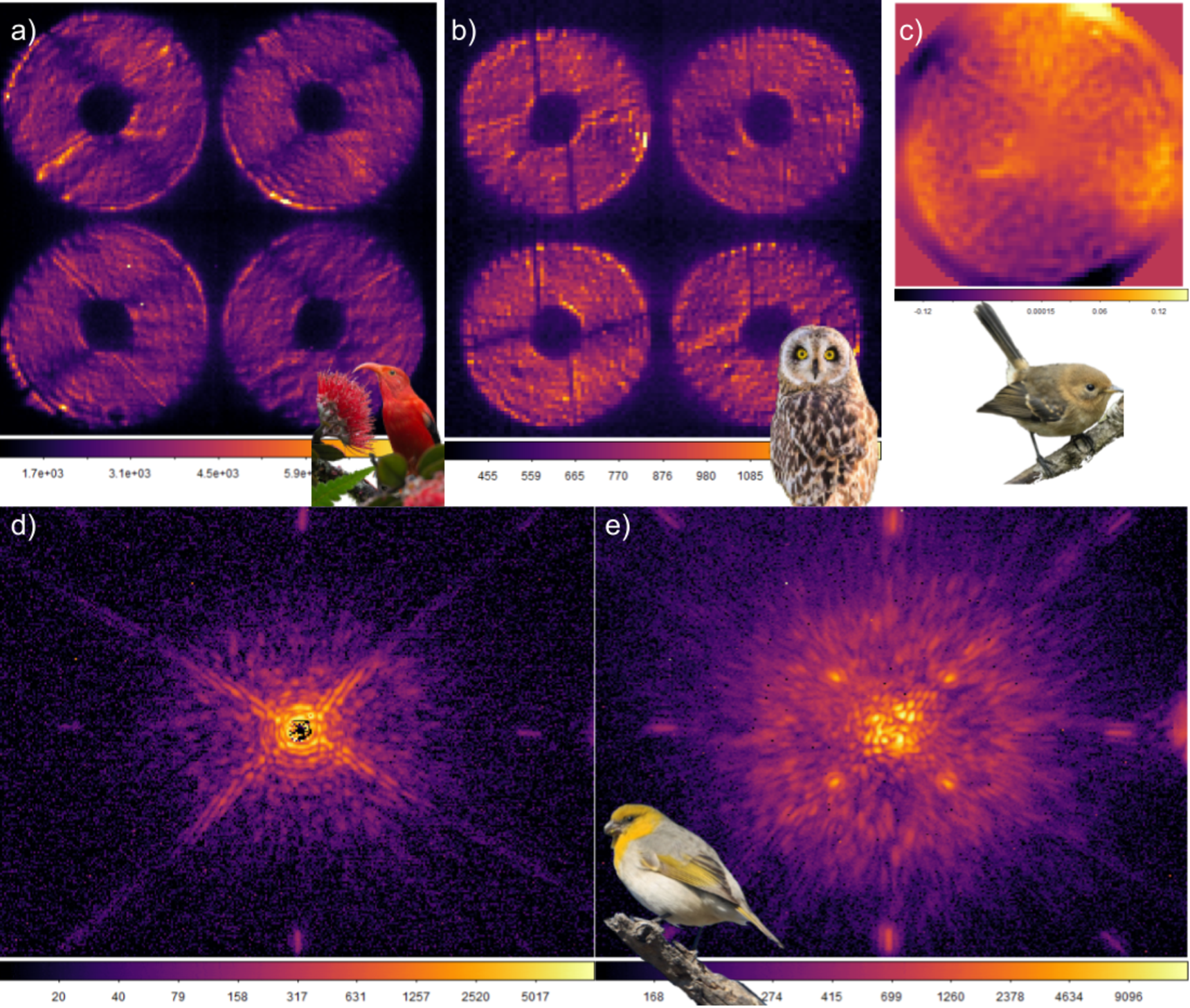}
\end{tabular}
\end{center}
\caption{Still frames during the first on-sky closed loop demonstration of two XAO loops running at the same time: AO3k (with the NIR WFS @ 1~kHz and ALPAO's DM3k) and SCExAO (with the visible PyWFS and BMC's MEMS 2k DM), on the target H\=ok\=ule`a (Arcturus). (a) AO3k's PyWFS image (`I`iwi) in H-band. (b) SCExAO's PyWFS image (Pueo) in I-band. (c) DM3k command applied by the NIR WFS (`Elepaio). (d) Saturated H-band PSF on SCExAO's Palila. The control region of the SCExAO loop (diameter 45~$\lambda/D$) is visible. (e) Coronagraphic PSF using a Lyot coronagraph. The control region is again visible, as well as the 25~nm astrometric speckle grid at 15~$\lambda/D$.}
\label{fig:ao3k_allcams} 
\end{figure} 

Figure~\ref{fig:ao3k_allcams} presents still frames of the various streams that were recorded during the night with CACAO, on the target H\=ok\=ule`a (Arcturus), the "star of gladness", used by Polynesian navigators as the zenith star indicating the latitude of Hawai`i:
\begin{itemize}
    \item (a) The PyWFS mode of the NIR WFS using the C-RED ONE camera and its mascot the `I`iwi, a vulnerable endemic species of Hawaiian honeycreeper. It uses 50\% of the H-band light in this example.
    \item (b) The PyWFS mode of SCExAO using the OCAM2k and its mascot the Pueo, the endangered species of Hawaiian Owls.
    \item (c) The DM command computed by the new RTC and sent to the DM3k, with its mascot the Hawai`i `Elepaio, an endangered species of monarch flycatcher.
    \item (d) and (e) Respectively a saturated image and coronagraphic image of the H-band PSF with the Palila camera (C-RED 2). In both images, the "dark hole" or control region of the SCExAO DM, with a diameter of 45~$\lambda/D$ is visible. The coronagraph used in (e) is a Lyot coronagraph with an inner working angle of 116~mas. An astrometric grid of speckles at 15.9~$\lambda/D$ is also visible here.
\end{itemize}

Not pictured here is the command sent to the SCExAO DM, and its mascot the N\=en\=e, a vulnerable species of Hawaiian goose and Hawai`i's state bird.

Pretty much right away, we demonstrated excellent corrections with both loops, although some improvements need to be addressed. The AO3k loop was correcting 2000~modes, while the SCExAO loop was correcting 1200~modes. We can potentially increase these numbers a little. Also, the NIR PyWFS would lock into petal modes that split the PSF into two to four lobes, depending on the profile of the modal gain used. Petal modes are known to be difficult for PyWFSs to control, but we are working on several options to address this issue.

\begin{figure} [b]
\begin{center}
\begin{tabular}{c}
\includegraphics[width=0.97\textwidth]{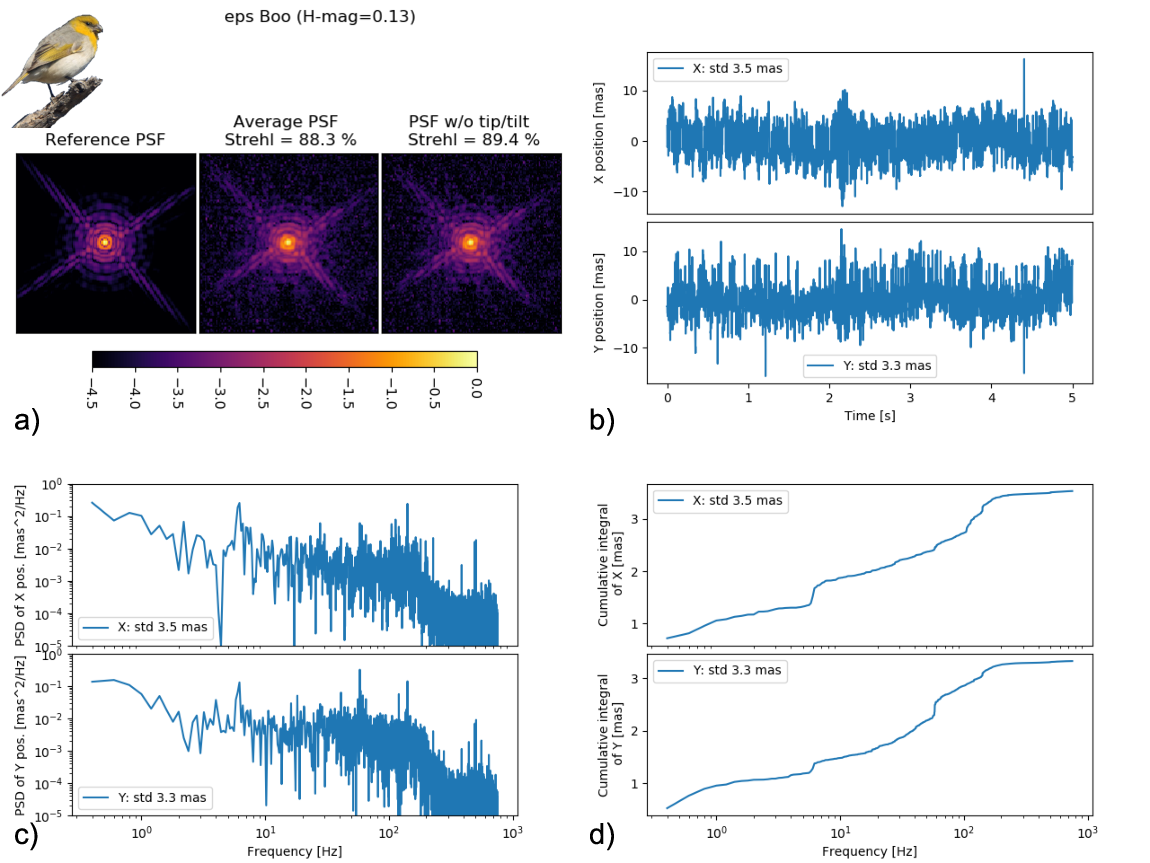}
\end{tabular}
\end{center}
\caption{Analysis of tip/tilt residuals using 5 seconds of unsaturated Palila data (1.5~kHz). (a) Comparison of the on-sky PSF with the reference PSF used to calculate the Strehl ratio, with an averaged PSF and a tip/tilt corrected averaged PSF. (b) temporal measurement of the position of the PSF over the sequence. (c) PSD of the measured position. (d) Cumulative integral of the PSD of each axis. The steps show which vibrations are the most significant. }
\label{fig:ao3k_tt} 
\end{figure} 

Figure~\ref{fig:ao3k_tt} presents the analysis of a 5~s sequence of unsaturated images taken with Palila at 1.5~kHz, on the star $\epsilon$ Boo. Figure~\ref{fig:ao3k_tt} (a) shows the reference PSF computed for the Strehl calculation, the simple temporal average of the data cube, and the average after shifting each frame to correct for tip/tilt residuals. The Strehl ratio go from 88.3\% to 89.4\%, showing a marginal improvement of only 1\%. The final PSF is mostly dominated by some small low-order non-common path errors that we could not offset that night. On Fig.~\ref{fig:ao3k_tt} (b), we can see the temporal measurement of the PSF position in X and Y. The RMS residual is only 3.5 and 3.3~mas for each axis respectively. We saw residual tip/tilt between 2 and 4~mas on that night, both in IR and in visible with VAMPIRES. Figure~\ref{fig:ao3k_tt} (c) shows the power spectrum density (PSD) of the same data. Mechanical resonances are seen on both axes on a wide range of frequencies, especially around 5~Hz, 60~Hz, 120~Hz and 500~Hz. The peak around 5~Hz is known to originate from the telescope, while the other ones are probably coming from the instrument itself. Figure~\ref{fig:ao3k_tt} (d) shows the cumulative integral of the PSD on both axes, telling us which vibrations contribute the most to the measured residual: the higher the step, the stronger the vibration is. In this case, vibrations are not a huge contributor, but we can still notice the telescope vibration around 5~Hz, and the 60~Hz vibration. A predictive control can be used to help correct for these, and improve even more the tip/tilt stability.

More on-sky time is needed to characterize the double-XAO correction, especially the contrast stability in coronagraphic mode, that most of SCExAO's users are interested in. 

\subsection{AO3k Providing High-Strehl With the NIR WFS}
\label{sec:ao3kmode}

During the engineering nights, we also took time to characterize the AO3k system by itself, using SCExAO passively to record the telemetry. The integral field spectrograph CHARIS was also used to record the PSF in the range 1.1 to 2.4~\mum (J, H and K-band). 

\begin{figure} [b]
\begin{center}
\begin{tabular}{c}
\includegraphics[width=0.97\textwidth]{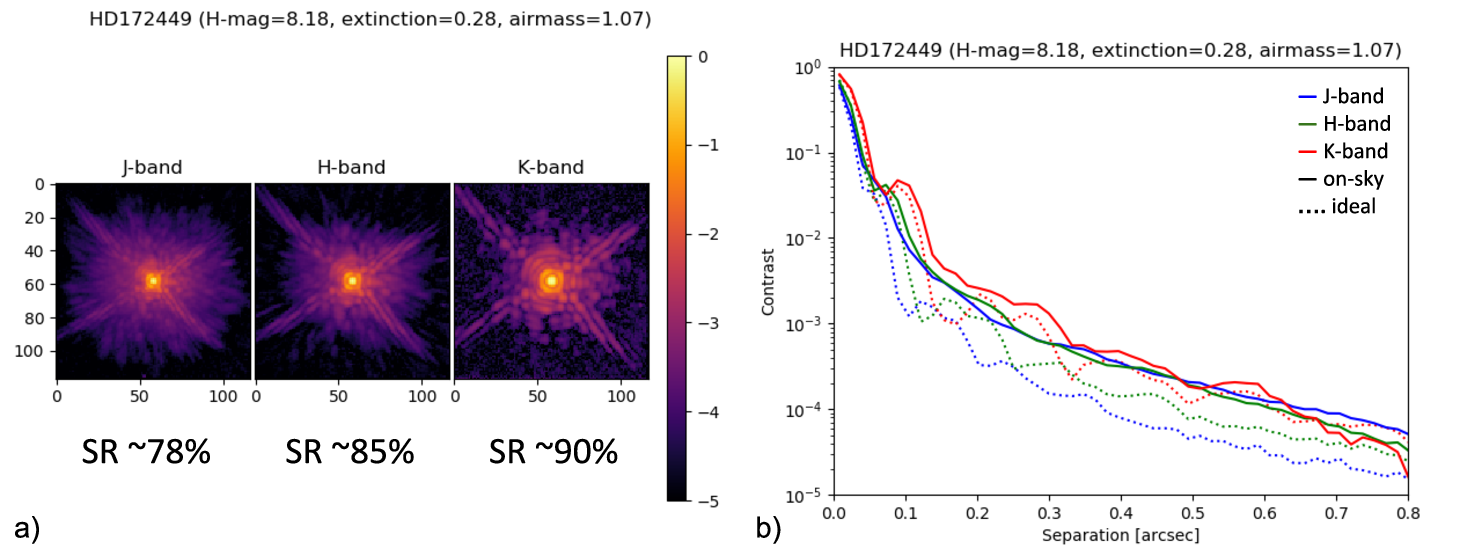}
\end{tabular}
\end{center}
\caption{(a) Strehl measurement on the target HD 172449 using CHARIS, in J, H and K-band, by co-adding the slices of the spectro-imager over each band. (b) Radial profile of the three PSFs compared to simulated ideal PSFs.}
\label{fig:ao3k_charis} 
\end{figure} 

Even with AO3k alone, we achieved excellent Strehl ratios at all wavelengths, like on the example presented in Fig.~\ref{fig:ao3k_charis} (a). Here, the slices  were co-added to create a image for each band. As expected, the Strehl ratio gets better at longer wavelengths, as the speckle halo and low-order errors are less pronounced. Figure~\ref{fig:ao3k_charis} (b) shows the radial profile of each band (solid lines), compared to an ideal profile (dotted lines). The impact of the residual speckle halo is noticeable especially in J and H-band, after 0.2~arcsec separations.

\begin{figure}
\begin{center}
\begin{tabular}{c}
\includegraphics[width=0.97\textwidth]{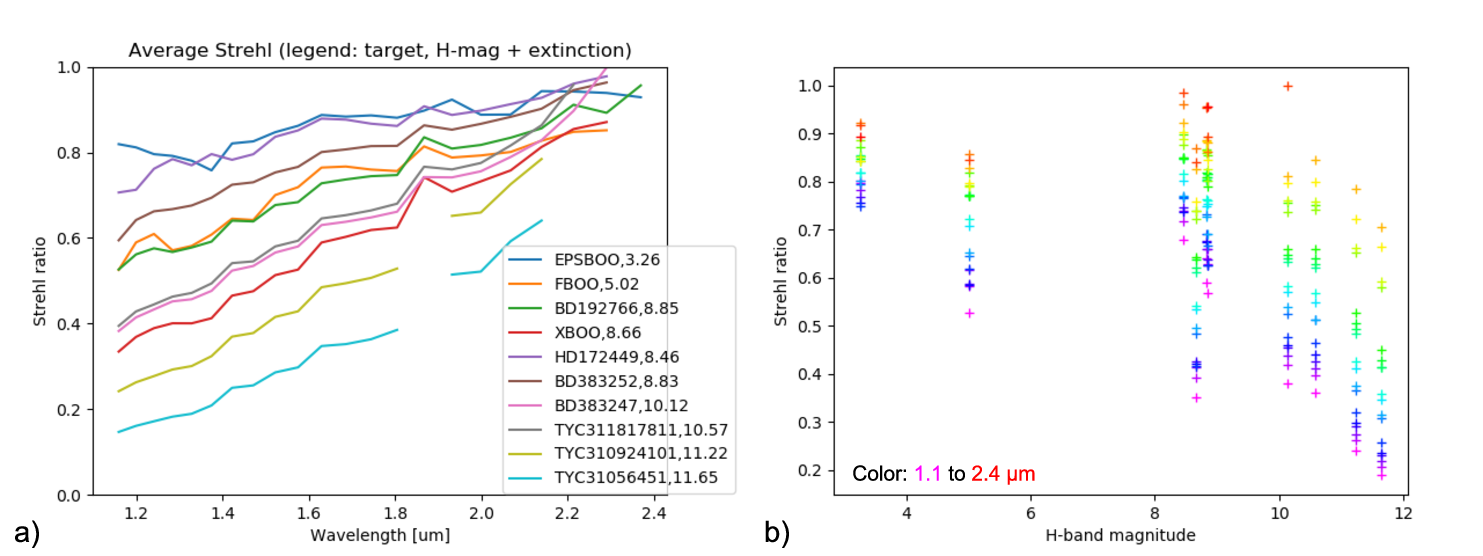}
\end{tabular}
\end{center}
\caption{(a) Strehl measurements between 1.1 and 2.4~\mum using CHARIS, on several targets with various magnitudes. (b) The same measurements displayed as Strehl versus H-band magnitude (corrected for extinction due to clouds). The performance starts to decrease after magnitudes 9-10 in this case.}
\label{fig:ao3k_strehlvmag} 
\end{figure} 

We observed a wide variety of star magnitudes, between 0 and 12th magnitudes in H-band. The Strehl measurements for each star are presented in Fig.~\ref{fig:ao3k_strehlvmag} (a). Figure~\ref{fig:ao3k_strehlvmag} (b) presents the same results, but as Strehl versus mH-band magnitude (taking into account the extinction from high-altitude clouds we had during some of the measurements). Similarly to what we saw in Fig.~\ref{fig:srmeasurements} (b), the Strehl ratio is high up to magnitude ~9, then starts to go down as we get less photons in the NIR WFS. But even close to the magnitude limit of 12, the loop was still providing some decent correction.

These first results show that, with the NIR WFS, AO3k provides very stable high-quality PSFs already, thanks to the high number of actuators and the high loop speed. We should expect similar results with the nlCWFS in the near future, which will then complete phase Ib of the upgrade of AO188 into AO3k.

Since the DM upgrade was successful, the science programs already expected to use the NIR WFS will see a significant improvement in the data quality going forward. Most notably, we looked at the galactic center with IRCS and produced the highest quality data ever recorded at Subaru.

\section{CONCLUSION AND PERSPECTIVE}

With this work, we demonstrated that the well deserved upgrades of AO188 into AO3k were successful. Despite plenty of delays, we managed to replace the DM with ALPAO's DM3k in time for our scheduled on-sky demonstrations, without compromising on the capabilities of the facility AO, and without impacting the scheduled Open-Use observations.

Combined with the NIR WFS installed during Phase Ia of the scheduled upgrades, the new DM3k provided the most stable PSFs for a facility adaptive optics on an 8-m class telescope. In addition, we corrected the atmospheric turbulence with two XAO systems simultaneously, AO3k and SCExAO, producing very stable high-contrast images with tip/tilt residuals down to a few milliarcseconds level. This will open new capabilities, with for example on-sky dark digging, or in combination with innovative astrophotonics technologies. Even more classical all-purpose instruments like IRCS will benefit from AO3k, with better and deeper image quality, or higher spectroscopic resolutions. Phase Ib will be fully complete this year with the addition of the nlCWFS, which should provide similar correction than the NIR WFS, without having to share the NIR science photons. With the addition of the NBS early next year, we will have the most complete and versatile AO platform of any telescope.

The next few years are truly going to be exciting for high-contrast imaging at the Subaru Telescope, as we will be able to reach deeper and more stable contrasts, around much redder targets. The technologies tested now will also be essential for the next generation of telescopes like TMT.

\acknowledgments 
Based on data collected at Subaru Telescope, which is operated by the National Astronomical Observatory of Japan. The development of AO3k is supported by the Japan Society for the Promotion of Science (Grant-in-Aid for Research \#21H04998, \#19H00695 and \#17H06129), the Subaru Telescope, the National Astronomical Observatory of Japan, the Astrobiology Center of the National Institutes of Natural Sciences, Japan, and the Heising-Simons Foundation. The development of SCExAO is supported by the Japan Society for the Promotion of Science (Grant-in-Aid for Research \#23340051, \#26220704, \#23103002, \#19H00703, \#19H00695 and \#21H04998), the Subaru Telescope, the National Astronomical Observatory of Japan, the Astrobiology Center of the National Institutes of Natural Sciences, Japan, the Mt Cuba Foundation and the Heising-Simons Foundation. Maunakea is a mountain that connects Kanaka Maoli (Native Hawaiians) to their ancestral and universal origins. We, as scientists and educators, acknowledge this interconnectedness between Kanaka and their `\=Aina (land).  We recognize that we have a kuleana (responsibility) that comes with the opportunity to pursue astronomical research on Maunakea. It is our kuleana to build and strengthen healthy relationships with the land and the people of this place based on mutuality and trust. J.L. would like to thank Leinani and Makali`i for allowing him to write this proceeding between feedings, burping and diaper changes.


\end{document}